\documentclass[11pt,useAMS,usenatbib]{article}
\usepackage{fullpage}
\usepackage{graphicx}
\usepackage{fancyhdr}
\title{{\sc LARGE--SCALE EFFECTS OF IONIZING FEEDBACK}\\
Invited contribution to the Sexten Cluster Workshop\\`The Formation and Early Evolution of Stellar Clusters'\\23--27 July 2012, Sexten, Italy}
\author{James E. Dale\\Excellence Cluster `Universe', Garching, Germany\\
dale@usm.lmu.de}
\date{}
\begin{document}
\maketitle

\section{\sc Introduction}
Modelling the influence of stellar feedback on Giant Molecular Clouds (GMCs) is plainly a very important task and there are many authors working on this issue, e.g. \cite{vs,cunningham, offner,gendelev}. Simulations which neglect feedback are very complex in their own right. Introducing additional physical effects, e.g. feedback mechanisms such as HII regions or winds, of course makes them even more so and runs the risk of making the calculations as difficult to understand as real star--forming regions. Due to this increased complexity, the best approach in my view is to introduce new physical effects one at a time and to compare simulations with control runs in which the new effect is absent. In reality, several feedback mechanisms act simultaneously and their effects are very unlikely to add together linearly, so that additional simulations in which multiple effects are included will be inevitably required. However, such simulations will be much easier to interpret if we know what each mechanism does when acting alone.\\
\indent Here I discuss the action of expanding HII regions driven by multiple O--type stars in GMCs. Full details of my approach can be found in \cite{deb2012a}. In this contribution, I highlight several ideas about HII regions which certainly were commonly held in the past, and which are often still espoused today, that I believe recent simulations by myself and others cast some doubt upon:\\ 
(i) HII regions are able to expel large quantities of gas from GMCs\\
(ii) HII regions expel gas from embedded clusters on timescales short enough to be dynamically important for their stellar component\\
(iii) HII regions raise the star formation efficiencies in clouds by triggering\\
(iv) Feedback from HII regions synchronises star formation and reduces stellar age spreads\\
\section{\sc Numerical simulations}
\begin{figure*}
\includegraphics[width=\textwidth]{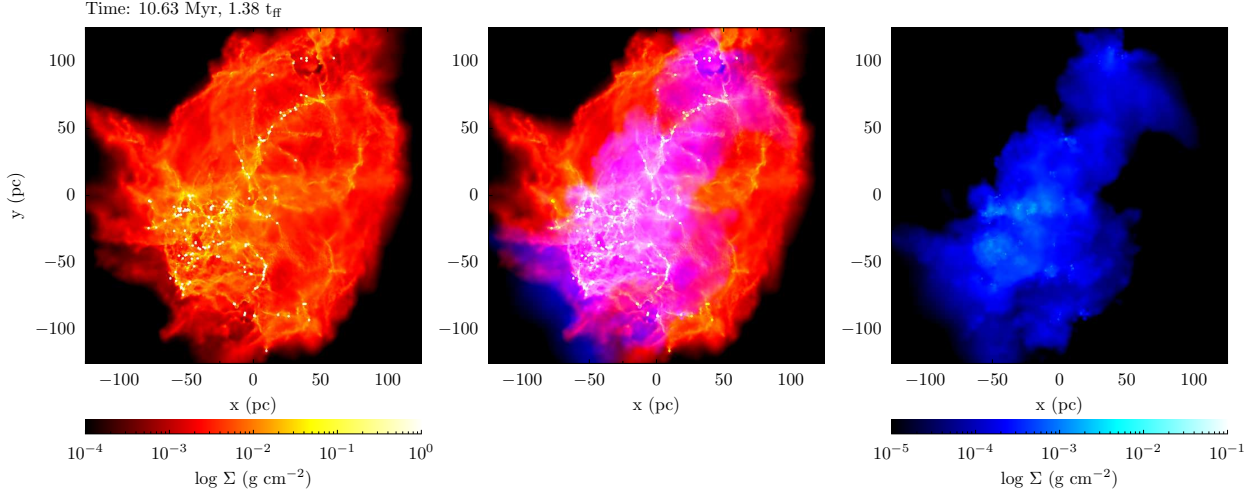}
\caption{Column density maps of all gas (left panel), HII gas only (right panel) and both neutral and ionized gas superimposed (centre panel) from Run B. Note the different column--density scales.} 
\label{fig:runb}
\end{figure*}

\begin{figure*}
\includegraphics[width=\textwidth]{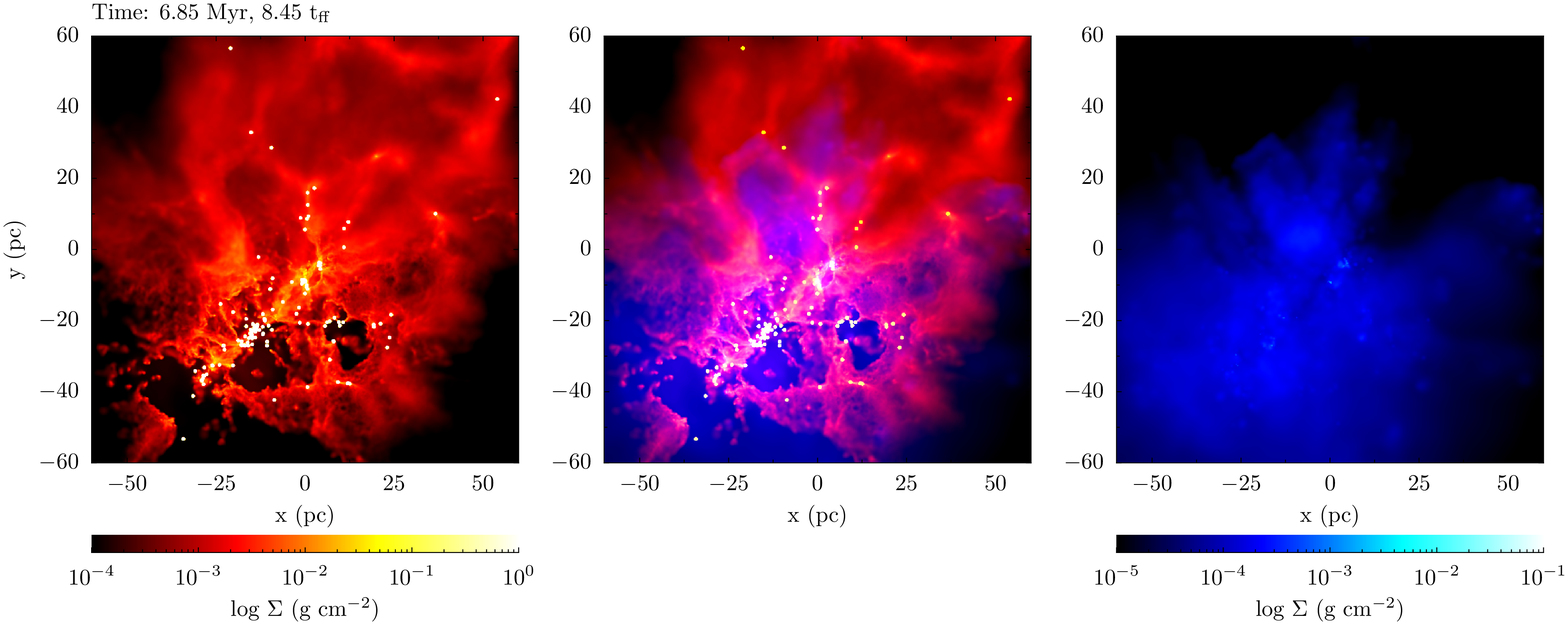}
\caption{Column density maps of all gas (left panel), HII gas only (right panel) and both neutral and ionized gas superimposed (centre panel) from Run UU. Note the different column--density scales.} 
\label{fig:runu}
\end{figure*}

\begin{figure*}
\includegraphics[width=\textwidth]{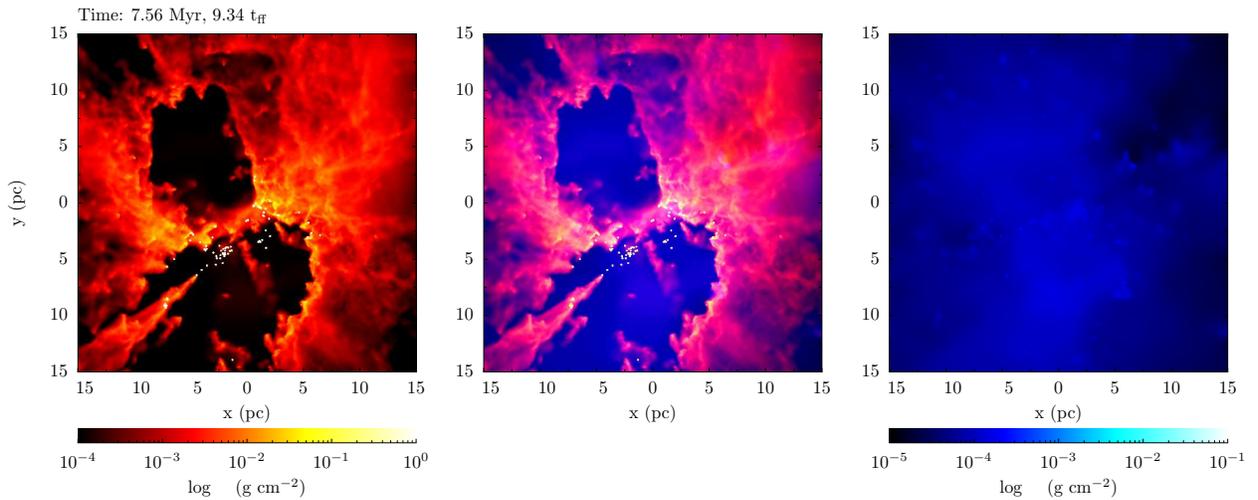}
\caption{Column density maps of all gas (left panel), HII gas only (right panel) and both neutral and ionized gas superimposed (centre panel) from Run I. Note the different column--density scales.} 
\label{fig:runi}
\end{figure*}

In a series of recent and forthcoming papers \cite{db2012,deb2012a, deb2012b, deb2012c}, I have been conducting a parameter study of the reaction to photoionizing feedback of turbulent star--forming clouds of a range of masses (10$^{4}$--10$^{6}$M$_{\odot}$), radii (2.5--180pc) and turbulent velocity dispersions (2.1--13.8 km s$^{-1}$). Simulations are performed using a version of the Smoothed Particle Hydrodynamics code described in \cite{benz90}. Stars or subclusters are represented by sink particles \cite{bate95} and ionizing radiation from the clouds' stellar component is simulated using a version of the photoionization code described in \cite{dec2007}, modified to allow for multiple ionizing sources as described in \cite{deb2012a}. Two batches of clouds have been simulated, one with a fixed initial virial ratio (defined here so that virial equilibrium is 0.5) of 0.7 and a second with a fixed ratio of 2.3. The action of ionization is followed for as near as possible 3 Myr (which we refer to as $t_{\rm SN}$) to determine its effectiveness when acting in the absence of stellar winds or jets, and before the detonation of any supernovae.\\
\indent In Figures \ref{fig:runb}, \ref{fig:runu} and \ref{fig:runi}, we show examples of these simulations. Run B, shown in Figure \ref{fig:runb}, is an initially bound 10$^{6}$M$_{\odot}$ cloud from \cite{deb2012a}. Ionization does little to the dynamics of this cloud, unbinding only a few percent of its gas reserves on the $t_{\rm SN}$ timescale and having only a very small effect on the star formation rate or efficiency. Run UU in Figure \ref{fig:runu} is an initially partially unbound 10$^{5}$M$_{\odot}$ cloud from \cite{deb2012c}. Although photoionization creates several bubble--like structures and a fine example of a champagne flow centred at about (-35,-40)pc, the dynamical effect of feedback is again modest, with an additional $\sim7$ percent of the gas becoming unbound (on top of that unbound from the beginning by the supervirial velocity field) and the star formation efficiency being reduced from 10 to 8 percent compared to the control simulation. Run I in Figure \ref{fig:runi} is an initially--bound 10$^{4}$M$_{\odot}$ cloud from \cite{deb2012a} on which ionization has a substantial effect, unbinding approximately 65$\%$ by mass of the cloud, although only $\sim12\%$ by number of the stars are unbound. The cloud structure is profoundly modified, sporting two large bubbles and a prominent pillar in the bottom left corner of the image.\\
\section{\sc Discussion and Conclusions}
\indent The principal results of the numerical study are:\\
\indent (i) The quantity of gas expelled by the HII regions over the fixed 3Myr timescale is very strongly dependent on the cloud escape velocity, regardless of whether the clouds are initially bound. Since clouds are observed to have a roughly fixed surface density (see \cite{heyer}), one may write
\begin{eqnarray}
M_{\rm cloud}=AR_{\rm cloud}^{2}
\label{eqn:MAR}
\end{eqnarray}
where $A$ is a constant with units of surface density and a value of $\approx0.03$ g cm$^{-2}$. The cloud escape velocity is given by
\begin{eqnarray}
v_{\rm ESC}=\sqrt{\frac{2GM_{\rm cloud}}{R_{\rm cloud}}}
\label{eqn:vesc}
\end{eqnarray}
Combining Equations \ref{eqn:MAR} and \ref{eqn:vesc} yields
\begin{eqnarray}
v_{\rm ESC}=\sqrt{2G(AM_{\rm cloud})^{\frac{1}{2}}}
\end{eqnarray}
The escape velocity is thus a weak function of the cloud mass, varying only as $M_{\rm cloud}^{\frac{1}{4}}$, but GMC masses range over many orders of magnitude and even over the few decades in mass which characterize most Galactic GMCs \cite{heyer}, the escape velocity varies by a factor of $\sim5$. This is significant largely because the ability of HII regions to expel material from gravitational potential wells is set by their fixed thermal sound speed of $\approx10$km s$^{-1}$. A similar result has recently been put forward by \cite{bressert}. We find therefore that, while low--mass (10$^{4}$M$_{\odot}$) clouds may lose several tens of percent of their mass over the 3Myr time window allowed, the more massive clouds are almost unaffected (note that this picture may be changed for the most massive clouds if radiation pressure were included \cite{krumholz,murray}.\\
\indent (ii) Even when ionization is able to expel substantial quantities of gas from a given cloud, the expulsion is too slow to unbind the majority of the embedded stars. This is all the more so in clouds which are initially bound, but also applies to the initially partially unbound systems.\\
\indent (iii) By comparison with control runs and using techniques from \cite{db2011}, we find that triggering, in the sense of forming stars that would not otherwise be born, does happen locally. However, in all cases studied, the  star formation rates are overall generally slower in the clouds suffering from feedback. The resultant star formation efficiencies are always lower, and the total numbers of stars can vary in either direction, as some star formation is aborted by feedback. The global effect of ionizing feedback in these simulations is, however, to suppress star formation. The ionizing sources are almost always found embedded in the densest star--forming gas and the action of feedback, if any, is mostly to destroy this gas. The sweeping up of low density material that would not otherwise form stars is a second--order effect and cannot compensate for the damage done to clouds' principal sites of starbirth.\\
\indent (iv) Structures such as pillars, champagne flows and bubbles all emerge naturally from these simulations. This is arguably not surprising, but it is nevertheless reassuring and, in my view, studying such serendipitously--formed structures is safer than setting out to form them deliberately from more carefully--constructed initial conditions. The very fact that randomly--seeded turbulent velocity fields offer so little control over the initial conditions and make the detailed outcome of simulations unpredictable is actually of value in preventing us from only finding the answers we were looking for. All these gaseous structures are commonly regarded as prime sites to look for triggered star formation. However, I find that the correlation of triggered stars with pillars and bubbles, while good, is by no means perfect. In particular, the stars located near the tip of the pillar seen in Figure \ref{fig:runi} are \emph{not} triggered. This pillar is in fact the photo--eroded remains of an accretion flow (or filament) which was feeding gas into the central cluster. In the absence of feedback, the material from which the stars at the pillar tip form is delivered to this cluster, where it is involved in star formation. Needless to say, this makes trying to infer which stars are triggered and which are not by observing a system at a single epoch exceedingly difficult.\\
\indent (v) Again, by comparison with simulations in which there is no feedback, we find that the effect of ionization is, if anything, to increase the spread in stellar ages. In the control simulations, the majority of stars or clusters continue accreting gas for the duration of the simulation, since there is nothing to stop them. In the feedback simulations by contrast, objects are often deprived of further supplies of cold gas either by the ionization front washing over them, or by themselves wandering into the HII regions. The times at which objects cease accreting in the feedback simulations are therefore largely stochastic, depending on where each star is with respect to the ionizing sources and the relative motion of each star with respect to the nearest ionization front.\\
\section{\sc Acknowledgements}
I wish to thank the Excellence Cluster `Universe' and the Ludwig-Maximilians-Universit\"{a}t M\"{u}nchen for financial support, and the organizers of the Sexten Cluster Workshop -- Nate Bastian, Rob Gutermuth and Linda Smith -- for inviting me to speak.\\

\end{document}